\documentclass[a4paper]{jpconf}

\usepackage{graphicx}

\usepackage{textcomp}           
\usepackage[latin1]{inputenc}   
\usepackage{wrapfig,rotating}
\usepackage{amssymb,amsmath,array}

\usepackage[]{lineno}

\newcommand{\unit}[1]{\ensuremath{\mathrm{\,#1}}}
\renewcommand{\u}[1]{\unit{#1}}
\newcommand{\um}[0]{\u{\mu m}}
\newcommand{\mm}[0]{\u{mm}}

\renewcommand{\th}{$^\textrm{th}$}

\begin{document}

\title{Test beam studies for a highly granular GRPC Semi-Digital HCAL} 

\author{Vincent Boudry, for the CALICE collaboration}   


\address{Laboratoire Leprince-Ringuet -  École polytechnique, CNRS/IN2P3\\
  91128 Palaiseau Cedex -- France}

\ead{Vincent.Boudry@in2p3.fr}

\begin{abstract}
  The Particle Flow Analysis approach retained for the future ILC detectors
  requires high granularity and compact particle energy deposition. %
  A Glass Resistive Plate Chamber based Semi-Digital calorimeter can offer both
  at a low price for the hadronic section.  %
  This paper presents some recent developments and results near test beam in
  the use of Glass Resistive Plate Chamber with embedded front-end electronics
  to build a prototype based on this principle. %
  All the critical parameters such as the spatial and angular uniformity of the
  response as well as the noise level have been measured on small chambers and
  found to be appropriate.  %
  Small semi-conductive chambers allowing for high rates and a large chamber
  have also been tested.
\end{abstract}


\section{Case for a GRPC Semi-Digital Gaseous Hadronic Calorimetry}
\label{sec:Case}

The Particle Flow Analysis (PFA) approach~\cite{Brient:2002gh}, retained for
all the detector concepts in development for the future International Linear
Collider (ILC), requires calorimetric systems able to spatially separate the
contributions from the individual particles constitutive of a jet.  %
The key feature of calorimeters then shifts from the best energetic resolution
in response to individual particle to the best spatial resolution.  %
This paradigm is pushed at its limit in the case of digital calorimetry, where
a single bit summarises the energy information for each cell.

A digital hadronic calorimetry has long been discussed~\cite{Ammosov:2002jq,
  Ammosov:2004gb}. %
Some recent GEANT4~\cite{GEANT4_1, GEANT4_2} based studies on a
Scintillator-Lead calorimeter~\cite{Matsunaga:2007zz} have shown the
feasibility, using 3 thresholds (``two bits'') and cell lateral sizes of
$1×1\u{cm²}$, to reconstruct the energy of single charged pions, up to
100\u{GeV}, with an energy resolution comparable to the one obtained using an
analog readout (``many bits'').  %

The use of a gaseous, hydrogen poor, planar sensitive medium presents many
advantages for a digital calorimetry. %
The main ones are:
\begin{itemize}
\item their insensitiveness to the dispersed, delayed and usually relatively
  badly measured neutron component of hadronic showers, resulting in narrower
  shower: typically a 100\u{GeV} pion is expected to be contained at 99\u{\%}
  in a stainless-steel gas calorimeter of section
  $70×70\u{cm²}$~\cite{Brient:2004xu}.  %
\item the ease to adjust the readout pad size to very small value if desired;
\item more specifically for single Resistive Plate Chambers (RPC), the proven
  possibility to make large surface, robust and reliable sensors at very low
  cost.
\end{itemize}

Among the many available techniques usable for large surface gaseous sensors, 3
are now being studied in the CALICE collaboration: MicroMEsh GAseous Structure
(MicroMegas) 
and Gas Electron Multiplier (GEM) 
in which the ionisation electron signal is amplified by a mesh after a drift,
and GRPC (Glass Resistive Plate Chamber), in which the amplification occurs
immediately; this article focuses on the later.

The efficiency ($\epsilon$) and cell multiplicity ($\mu$) in response to
minimum ionising particles are the two critical parameters required for the
intercalibration of the channels of a (semi)digital calorimeter.  %
They should be supplemented by the understanding of the response to a full
hadronic shower, which might affect the sensor itself or the embedded
electronics.  %
Finally, although this should not be a problem at the ILC, the recovery time
for various spatial scales and intensity, have to be measured, as well as the
sensitivity to high magnetic fields.

Technologically, the use of RPC in a large hadronic calorimeter faces a few
challenges, which should be integrated as early as possible in the
conception. %
The first is the industrial production and handling of large surface, uniform
sensors at low cost.  %
To keep compact showers, the ratio of low density material to absorber should
minimised, requiring thin sensitive and readout layer.  %
Practically, the goal is to reach $6\mm$ all inclusive (RPC and electronics)
for $20\mm$ stainless steel absorber.  %

Such devices have been produced in various sizes and the present paper
describes the test setup in test beam and the results obtained: the aim is to
validate the whole embedded electronics readout system in beam conditions
(including the use of local data storage and train reconstruction), to study
the GRPC behaviour as a function of the applied HV and thresholds, of incoming
particle angle, position, rate and to validate the technological choices prior
to the construction of a full scale 1\u{m³} prototype~\cite{M3Talk}.

\section{Glass RPC prototypes with an embedded readout}
\label{sec:Protos}

\begin{wrapfigure}{r}{0.55\columnwidth}
  \vspace*{-10pt}
  \centering
  \includegraphics[width=0.55\columnwidth]{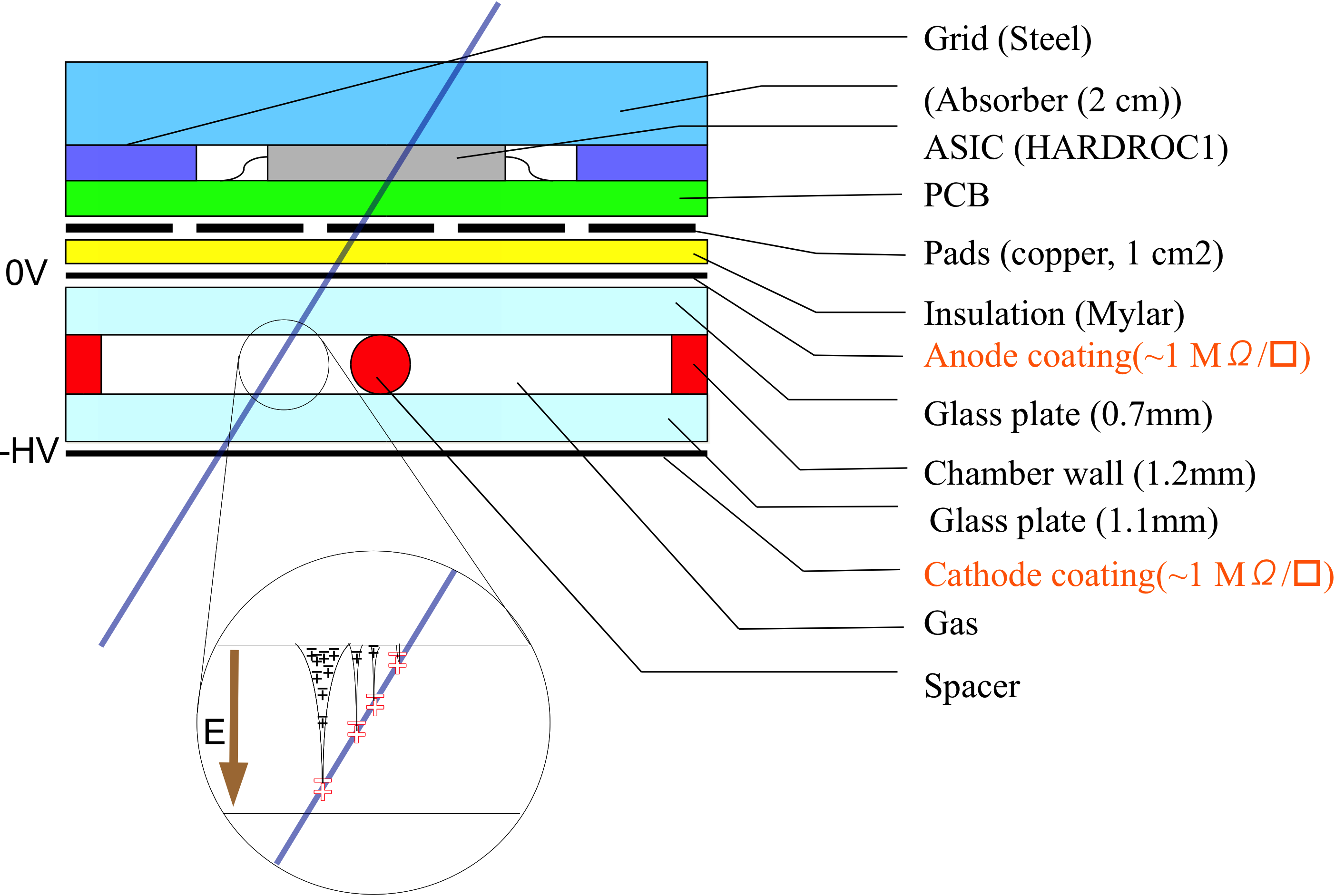}
  \caption{Transversal cut of the GRPC layout and readout electronics.}
  \label{fig:EvtDisp}
  \vspace*{-10pt}
\end{wrapfigure}

The results shown here were obtained with various prototypes sharing the same
geometry: %
a gap of $1.2\mm$ of gas between two glass plate, polarised under a nominal
voltage of $7.4\u{kV}$.  %
The gas used was a mixture 93\u{\%} of TetraFluoroEthane (TFE), 5\u{\%} of
Isobutane and 2\u{\%} of Sulphur HexaFluoride (SF6) yielding 8 primary
ionisations per mm of a mip track.  %
The glass plates, respectively of 1100 and 700\um\ thickness for the cathode
and anode sides, are separated by Nylon fishing lines and covered on their
external sides by a resistive coating.  %

Three types of coating were used: Graphite,
Statguard$^\textrm{\textregistered}$ and Licron$^\textrm{\textregistered}$
yielding resistivity of respectively 0.4, 2 and 20\u{M\Omega/\square}.

Small chambers of $8.35×33.55\u{cm²}$ were built using either standard float
glass, in combination with the 3 types of coating, or semi-conductive glass with
Licron and Statguard coatings.
Large chambers of $1×1\u{m²}$ were built out of float glass coated by colloidal
graphite (having resistivity of 1-2\u{M\Omega/\square}).


In order to read efficiently many channels, the very front-end part of the
readout electronics is integrated in the detector itself: the readout ASICs and
their supporting PCB, are only separated from the anode coating by a 50\um\
insulating layer of Mylar$^\textrm{\textregistered}$.


With a capacity of 64 channels, the HaRDROC~\cite{HardRoc1:2007} (Hadronic RPC
Detector Read Out Chip) specifically developed by the Omega 
micro-electronics group of Orsay, has 2 independent adjustable thresholds
(above $10\u{fC}$) used for auto-triggering.  %
It can store up to 127 events composed of an 8-bit ASIC ID (8 bits), a time
stamp (24 bits by steps of 200\u{ns}) and the hit map ($2×64$ bits). %
Each channel gain can be adjusted over a range $0$--$4$ on 6 bits.  %
Its internal inter-channel cross-talk has been measured to be below 2\u{\%}.


Some test boards holding 4 HaRDROC's and their control and readout logic, were
produced and tested in 2007.  %
They feature 8-layer, $800\um$ thick PCB with buried and blind vias, ensuring
cross-talk below $0.3\u{\%}$, and the readout of $8×32$ pads of $1×1\u{cm²}$,
separated by 0.5\u{mm} gaps.  %
First used to validate the new electronics and acquisition scheme, they then
served for the beam tests equipped with GRPC sensors, as described hereafter.

The relative gain of the individual channels of each ASIC were first adjusted
by levelling the threshold responses to $100\u{fC}$ injected charges.  %
A factor of 4 was gained on the built-in spread of the effective threshold,
lowering it to 2.5\u{fC} for a complete board.  %

For the $1×1\u{m²}$ prototype, a new set of readout electronics have been
designed: the test board functions have been split between so-called ASU
(Active Sensor Unit) boards holding 24 HaRDROC ASICs each, which can be
coupled together at their ends, and a readout and controller card named ``DIF''
(for Detector InterFace).  %
A $1×1\u{m²}$ chamber is readout by 6 ASU connected to 3 DIF's, for a total of
9216 channels.  %
The acquisition is performed through USB by a software developed in the XDAQ
framework. 

\section{Test beam setup}
\label{sec:MiniDHCAL}  

Up to 4 test boards were equipped with GRPC, of various types, and were used
in test beam near the PS and SPS CERN facilities.  %
They were installed perpendicularly to the beam in a structure eventually
allowing for the installation of 2\u{cm}-thick stainless steel radiators, to
build a mini calorimeter, associated with a set of scintillators for triggering
purposes.

Two campaigns of tests were performed in 2008 and 2009 each using low
($3$--$12\u{GeV}$) and high ($10$--$150\u{GeV}$) energy pions.  %
The results presented here are restricted to the mip-like data sample (no
absorbers used).

\subsection{Acquisition modes}
\label{sec:AcqModes}

The on-board FPGA's were programmed to manage two acquisition modes relying on
\begin{wrapfigure}[19]{r}{0.5\columnwidth}
  \centering
  \vspace*{-10pt}
  \includegraphics[width=0.5\columnwidth]{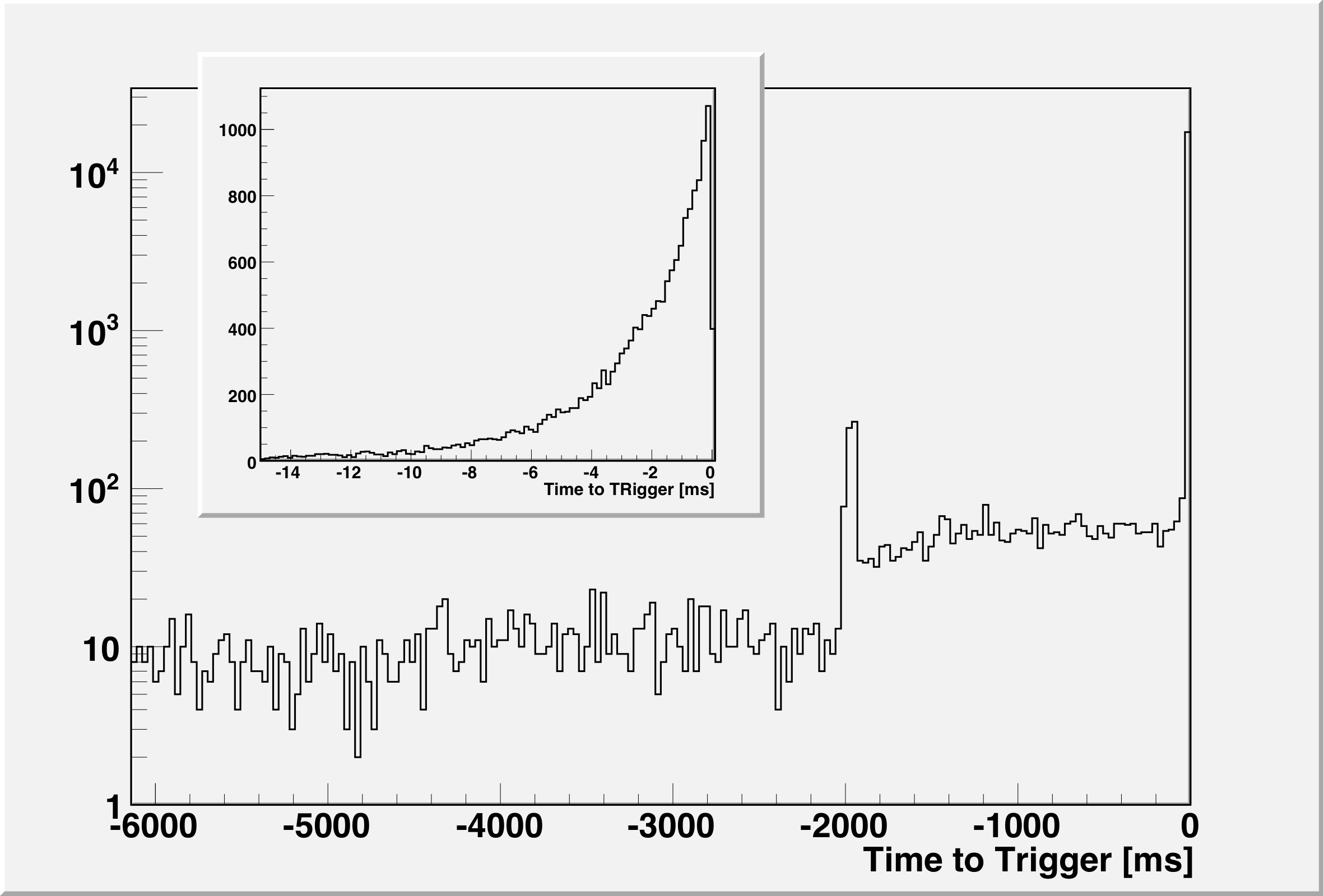}
  \caption{Time-to-trigger distribution; the events associated with the
    trigger are in the last 200\u{ns} (right peak),
    in-spill events in the last 20\u{ms} (detail in insert) while the previous
    spill is visible at -2000\u{ms}.  All events in between are considered as
    being noise. 
  }
  \label{fig:Timing}
\end{wrapfigure}
the auto-trigger feature of the HaRDROC's:
\begin{description}
\item[a train mode] (``ILC like''), where the readout is started on the
  start-of-spill signal and stopped on the end-of-spill signal or when one of
  the ASIC is full;
\item[a triggered mode] used for cosmic rays and test beam data taking, where
  the acquisition is stopped on an external trigger.  %
  When the memory one of the ASIC's is full, a busy signal is emitted, the
  entire board reset and the acquisition resumed.
\end{description}
In both cases when the acquisition is stopped, a train of up-to 127 events is
readout. %

Using the coincidence of plastic scintillator plates readout by
photo-multipliers as an external trigger a total of about 1,130 million
triggered events were collected, at an average DAQ rate of $\sim20\u{Hz}$.

\subsection{Timing}
\label{sec:Timing}

At PS, 2 spills of 400\u{ms} were received, at 2 second interval, every 48\u{s}
during the day (every 33\u{s} during the night).

The synchronisation between the boards was ensured by a $40\u{MHz}$ counter
recording the time difference for each board between the last internal trigger
and the external trigger.  %
This allows to resynchronise of each board w.r.t.\ the trigger signal and to
reconstruct the history of all boards back to the last memory reset.  %
In most cases (low noise and particle density), the last event in memory of the
ASIC in beam corresponds to the signal of the triggering particle.

By studying the correlation in time between hits of different cards, it has
been observed that their clock's dispersion is small enough to allow a tight
cut of $300\u{ns}$ (w.r.t.\ their 200\u{ns} period) to cluster hits in
individual events due to a particle or noise.

This time reconstruction allows the use of events from memory which were not
triggered by the scintillators in the beam.  %
Such a time reconstruction is shown in figure~\ref{fig:Timing}. 
In particular the noise pattern can be separated from in-spill events,
dominated by particles.

\section{Small chambers results}
\label{sec:SmallCh}

\subsection{Efficiency \& multiplicity studies}

\begin{wrapfigure}[20]{r}{0.5\columnwidth}
  \vspace*{-25pt}
  \centering
  \includegraphics[width=0.47\columnwidth]{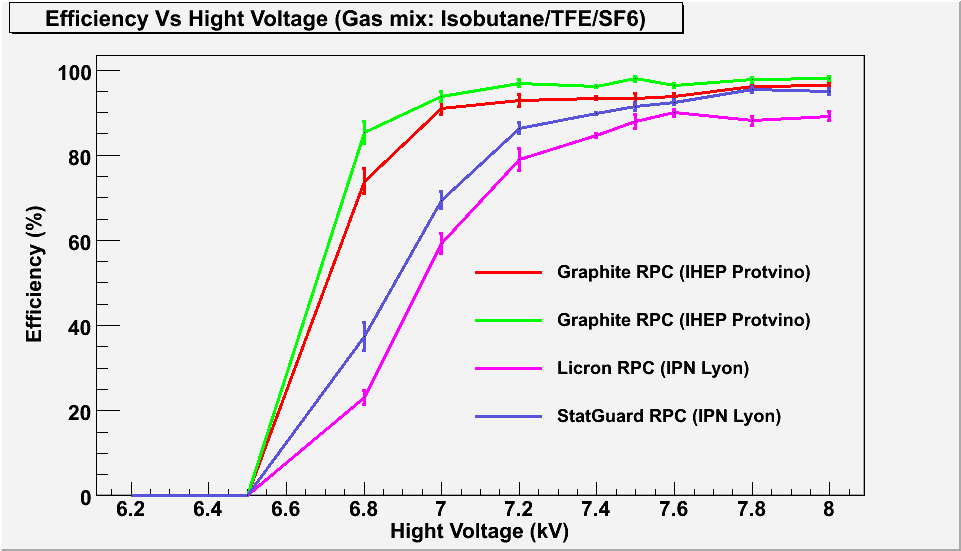}
  \includegraphics[width=0.47\columnwidth]{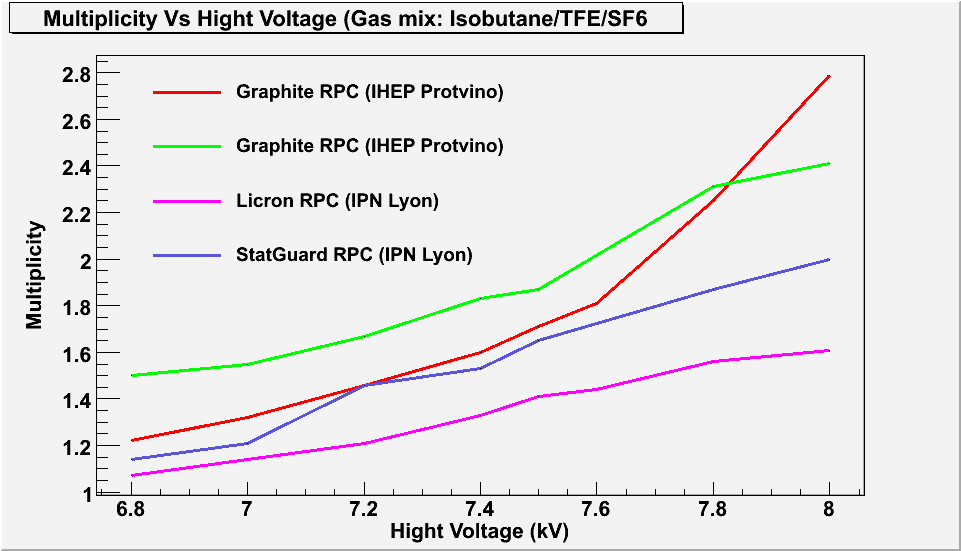}
  \caption{Efficiency and multiplicity vs HV for the 4 small float glass
    chambers.}
  \label{fig:BeamEffMult}
\end{wrapfigure}
Using the 3 other chambers as a crude tracking device one can determine the
position of the particle in the 4\th\ one and calculate the efficiency and
multiplicity locally.  
The multiplicity $\mu$ is defined as the number of fired cells within
3\u{cm}-radius around the impact when the chamber has responded ($\mu\ge1$)

For the following studies one of the two following selection was applied : %
\begin{itemize}
\item 
  using only the last event in memory (triggering one), with $t>-400\u{ns}$ %
\item 
  using the complete train statistics, in which case only the hits in time
  ($|\Delta t| < 300\u{ns}$ and after the most recent board reset ($t>$ time of
  first event in every chamber) were used.
\end{itemize}

\paragraph{A high voltage}
scan was performed with a threshold set to $165\u{fC}$ with the different small
chambers based on float glass (2 identical ``Graphite'', 1 ``Licron'' and 1
``Statguard'').  
Using the triggered event sample it was found that a maximum efficiency of
about 98\u{\%} was reached for all chambers but for the Licron one saturating
at $\sim90\u{\%}$.  %
As can be seen on figure~\ref{fig:BeamEffMult} the efficiency and multiplicity
were found to rise faster for the graphite covered chambers w.r.t.\ the Licron
and Statguard ones.  %
An efficiency plateau is reached for all chambers around $7.2\u{kV}$, while the
multiplicity rises continuously with the HV. %

A working point of $7.4\u{kV}$ (for a threshold of $165\u{fC}$) was chosen as
standard for the following studies, being seen as a good compromise between
high efficiency and low multiplicity.  %
At this point the multiplicities are of $1.3$ for the ``Licron'' chamber and
$\sim1.6$ for the ``Statguard'' and ``Graphite'' ones.

\paragraph{Threshold scan:}
The efficiencies and multiplicities were measured as the threshold applied was
varied around $165\u{fC}$, and found to behave as expected, as the inverse of
the HV; the efficiency was found to be $80\u{\%}$ at 7.4\u{kV} for a threshold
of 1.1\u{pC}.  %
This will be used to model the response of the chambers.

\paragraph{Chamber uniformity:}
The response ($\epsilon, \mu$) of 4 similar small graphite chambers produced by
the IHEP at Protvino were studied and found to have identical behaviour with a
spread of the efficiency of $95\pm3\u{\%}$ and multiplicity of $1.6\pm0.4$.

\begin{wrapfigure}[15]{r}{0.6\columnwidth}
  \vspace*{-10pt}
  \centering
  \includegraphics[width=0.6\columnwidth]{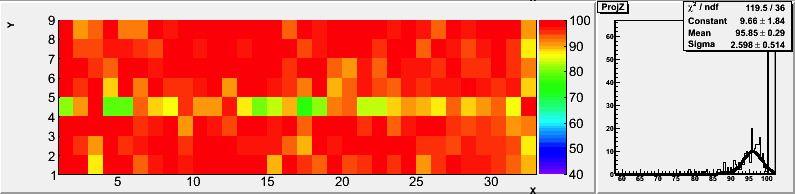}
  \includegraphics[width=0.6\columnwidth]{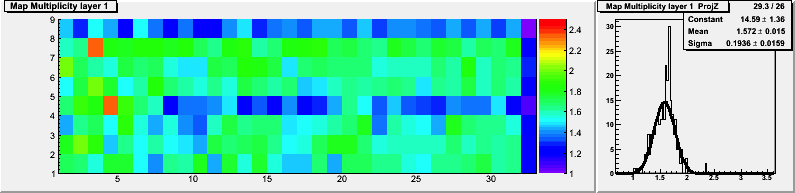}
  \caption{Efficiency (top) and multiplicity (bottom) spatial distribution on
    two chambers with 1\u{cm²} resolution (left) and corresponding values
    distribution (right). }
  \label{fig:BeamEffMult}
\end{wrapfigure}
Using the train sample reconstruction individual response of the $1×1\u{cm²}$
cells could be studied with a good statistics.  %
A global efficiency spread, including the statistics error for about 25k
events, of 3\u{\%} was observed inside single chambers and between different
chambers. %
The multiplicity spread in a chamber is of $\sim0.2$.  %
The main sources of variation (lower $\epsilon$ and $\mu$) are due to the
fishing lines and the chamber edges as can clearly be seen on
figure~\ref{fig:BeamEffMult}.  %
Some inefficiencies were also clearly observed at the centre of the beam spot
for high rates run (19\u{kHz/cm²}).

In the future prototypes, the fishing lines will be replaced by ceramic ball,
reducing the size of the affected region by a factor of 100.

\paragraph{Angular response \& stability:}
No observable difference was seen while shooting with an angle up to
60\u{^\circ}. %
This will ease the reconstruction code for large detectors.

The global behaviour was found to be constant over the time of the running (10
days), with a number of dead or inefficient cells limited to 1\u{\%}.

\subsection{Noise studies}
\label{sec:Noise}

Observing only out-of-spill events and out-of-trigger cosmic data, one has
access to the noise distribution.
The noise rate was studied and found to be higher near the special regions
(especially fishing lines and to a lesser extend borders), with rates typically
of 20\u{Hz/cm²} for our standard threshold (165\u{fC}) going down to 10 times
less when the threshold is raised at 750\u{fC}.  %

In the bulk, the noise rate was found to be of 0.1\u{Hz/cm²}, with no dependence
on the applied threshold (between 165 and 750\u{fC}).  %
It evolves slowly with HV with a slope roughly estimated $\sim0.3\u{Hz/cm²/kV}$
around our working point.

\subsection{Semi-conductive GRPC's}
\label{sec:SemiCond}

Two small chambers provided by Tsinghua University and based on semi-conductive
glass ($10^{10}\u{\Omega/cm}$ against $10^{13}\u{\Omega/cm}$ for classical
float glass) have been tested at high rates.  %
A ``thin'' version features glass thickness of 0.83\u{mm} on the anode side and
1.1\u{mm} on the cathode one, coated with Statguard.  %
A ``thick'' version uses 1.1\u{mm} glass on both sides, with a Licron coating.

\begin{wrapfigure}[17]{r}{0.5\columnwidth}
  \centering
  \vspace*{-20pt}
  \includegraphics[width=0.45\columnwidth]{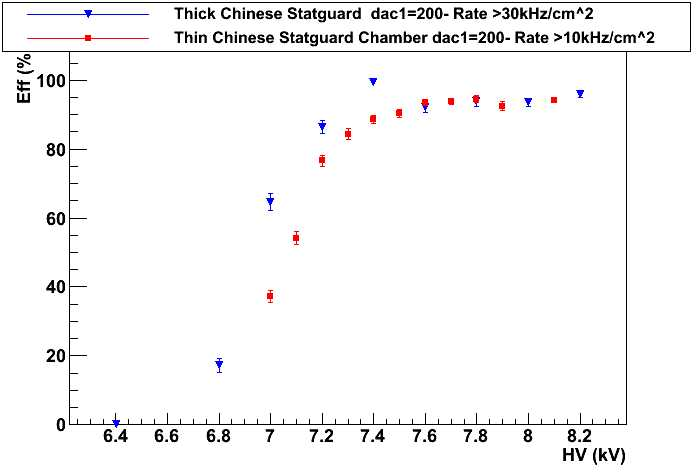}
  \caption{Efficiency of thick (blue triangles) and thin (red squares)
    semi-conductive glass at respectively average rates of
    $\gtrsim30$ and $\gtrsim10\u{kHz/cm^2}$, a threshold of 165\u{fC} and a
    voltage of 7.4\u{kV}.}
  \label{fig:SemiCond}
\end{wrapfigure}

As can be seen on figure~\ref{fig:SemiCond} an efficiency curve similar to the
standard chambers could be achieved, with efficiency reaching a plateau at
90\u{\%} above 7.4\u{kV}, but here with rates above 10 to 30\u{kHz/cm^2}, when
the standard chambers start to become inefficient above 10--100\u{Hz}.

\section{First results on a 1 m$^2$  prototype}
\label{sec:m2}

The square meter prototype was put in beam at the SPS.  %
The acquisition worked as expected and was able to read the 9216 channels; the
beam spot was observed and efficiencies of up to 93\u{\%} have been
obtained.  %
But the setup rapidly suffered from faulty mechanics on the HV connection and
these results will have to be confirmed.

On a cosmic test bench, using the mini-DHCAL as a tracker device, cosmic tracks
were collected.  %
For analysis purpose, the surface was divided as a checker of 9 regions of
$33×33\u{cm²}$.  %
One of the regions failed during this test.  %
Between the 8 other working regions a good uniformity of response was observed
with results similar to the ones of the small chambers (working HV
$\sim7.4\u{kV}$, $\epsilon\sim95\u{\%}$ and $\mu\sim1.6$).

\section*{Conclusions \& perspectives }

A Semi-Digital Gas Hadronic Calorimeter with embedded readout is a very
promising candidate for future linear collider experiments.  %
The main critical parameters have been checked at beam tests on float glass
small chambers prototypes: in response to single minimum ionising particle a
high efficiency ($\epsilon \sim 95\u{\%}$), low multiplicity ($\mu\sim1.6$), no
angular dependency, spatial and detector uniformity ($\sigma_\epsilon \lesssim
2\u{\%}$, $\sigma_\mu \lesssim 3\%$ over 256 cells of 1\u{cm²}) and low noise
have been measured.  %
Some prototypes using semi-conductive glass showed some promising performances
for high rates ($\gtrsim10\u{kHz}$).  %
The first measures made on a square meter sensor similar to the one foreseen
for a technological prototype of 1\u{m^3} (including 40 of these) were found to
be similarly satisfying. %
The full prototype is expected to be completed in 2010.

\section*{Acknowledgements}

The work on the electronics and the DAQ is supported by the Commission of the
European Communities under the 6$^\textrm{th}$ Framework Programme Structuring
the European Research Area, contract number RII3-026126.

\section*{References}
\bibliography{DHCAL_TB1_v0_arXiv}

\end{document}